\documentclass[prodmode,acmtecs]{acmsmall} % Aptara syntax

\usepackage[ruled]{algorithm2e}

\SetAlFnt{\small}
\SetAlCapFnt{\small}
\SetAlCapNameFnt{\small}
\SetAlCapHSkip{0pt}
\IncMargin{-\parindent}

\acmVolume{0}
\acmNumber{0}
\acmArticle{0}
\acmYear{2015?}
\acmMonth{0}

\usepackage{amsmath}
\usepackage{pgf,tikz,color}
\usepackage{pgfplots}
\usetikzlibrary{arrows}

\begin{document}

% Page heads
\markboth{B.~Ong et al.}{RIDC Methods: Software for Parallel-Time Integration}

% Title portion
\title{Algorithm xxx: RIDC Methods -- A Family of Parallel Time-Integrators}
\author{Benjamin W.~Ong
\affil{Michigan Technological University}
Ronald D.~Haynes
\affil{Memorial University of Newfoundland}
Kyle Ladd
\affil{Barracuda Networks}}

\begin{abstract}
Revisionist integral deferred correction (RIDC) methods are a
family of parallel--in--time methods to solve systems of initial
values problems.  The approach is able to bootstrap lower order time
integrators to provide high order approximations in approximately the
same wall clock time, hence providing a multiplicative increase in the
number of compute cores utilized.  Here we provide a library
which automatically produces a parallel--in--time solution of a system
of initial value problems given user supplied code for the right hand
side of the system and a sequential code for a first-order time
step.  The user supplied time step routine may be explicit or implicit
and may make use of any auxiliary libraries which take care of the
solution of any nonlinear algebraic systems which may arise or the
numerical linear algebra required.
%The code contains six examples of
%increasing complexity which also serve as templates to solve user defined
%problems.
\end{abstract}

\category{G.1.0}{Numerical Analysis}{Parallel Algorithms}
\category{G.1.7}{Numerical Analysis}{Ordinary Differential Equations
  -- Initial Value Problems}

\terms{Algorithms, Performance}

\keywords{Parallel--in--time, deferred correction}

\acmformat{Benjamin W.~Ong, Ronald D.~Haynes and  Kyle Ladd, 2014.
RIDC Algorithms: A Family of Parallel Time-Integrators}
% At a minimum you need to supply the author names, year and a title.
% IMPORTANT:
% Full first names whenever they are known, surname last, followed by a period.
% In the case of two authors, 'and' is placed between them.
% In the case of three or more authors, the serial comma is used, that is, all author names
% except the last one but including the penultimate author's name are followed by a comma,
% and then 'and' is placed before the final author's name.
% If only first and middle initials are known, then each initial
% is followed by a period and they are separated by a space.
% The remaining information (journal title, volume, article number, date, etc.) is 'auto-generated'.

\begin{bottomstuff}
This work is supported by AFOSR Grant FA9550-12-1-0455, and NSERC
Discovery grant (Canada).

Author's addresses: R.~ D.~Haynes, Department of Mathematics and Statistics,
Memorial University of Newfoundland; B.~W.~Ong,
Mathematical Sciences, Michigan Technological University; K.~Ladd, Barracuda Networks
\end{bottomstuff}

\maketitle

\section{Introduction}

The fast, accurate solution of an initial-value problem (IVP) of the form 
\begin{align}
  %{\bf y}'(t) = f_S(t,{\bf y}) + f_N(t,{\bf y}), \quad {\bf y}(0) = {\bf y}_0,
  {\bf y}'(t) = f(t,{\bf y}), \quad {\bf y}(0) = {\bf y}_0,
  \quad t\in[0,T],
  \label{eqn:ivp}
\end{align}
where ${\bf y}(t)\in \mathbb{C}^{N}$, 
$f:[\mathbb{R}\times  C^{N}]\to\mathbb{C}^N$, 
% $f_S:[\mathbb{R}\times C^{N}]\to\mathbb{C}^N$,
%  $f_N:[\mathbb{R}\times C^{N}]\to\mathbb{C}^N$,
 is of practical interest in scientific
computing.
%; Often, $f_S$ is a stiff relaxation term that needs to be
%discretized implicitly, and $f_N$ is a non-stiff advective or
%convective term.  
IVP~\eqref{eqn:ivp} often arises from the spatial discretization of
partial differential equations, and may require either an explicit or
implicit time-integrator.  The purpose of this software is to ``wrap''
a user-implemented first-order explicit or implicit solver for
IVP~\eqref{eqn:ivp} into a high-order parallel solver; that is, given
$(t_n,y_n,f_n)$, a user specifies a function that returns
$(t_{n+1},y_{n+1},f_{n+1})$ using either a forward Euler or backward
Euler integrator.  This work differs from existing ODE integration
software or libraries, where a user typically only needs to specify
the system of ODEs and relevant problem parameters.  The upside is
that our software provides a parallel--in--time solution while giving
the user complete control of the first-order time step routine.  For
example, the user may chose their own quality libraries for the
solution of systems of nonlinear algebraic equations or efficient
linear system solvers particularly tuned to the structure of their
problems.

There are three general approaches for a time-parallel solution of
IVPs \cite{burrage97}.  One approach is ``{\em
  parallelism-across-the-problem}'', where a problem is decomposed
into sub-problems that can be computed in parallel, and an iterative
procedure is used to couple the sub-problems.  Examples of this class
of methods include parallel wave-form relaxation methods
\cite{vandewalle89}.  The second approach is ``{\em
  parallel-across-the-step}'' methods, where the time domain is
partitioned into smaller temporal subdomains which are solved
simultaneously.  Examples of this class of methods include parareal
methods \cite{lionsmadayturinici01,gander07}, where the method
alternates between applying a coarse sequential solver and a fine
parallel solver.  The third approach is ``{\em
  parallelism-across-the-method}'', where one exploits concurrent
function evaluations within a step to generate a parallel time
integrator.  This approach typically allows for {\em small-scale}
parallelism, constrained by the number of function evaluations that
can evaluated in parallel.  This is often related to the order of the
approximation.  Examples of Runge--Kutta methods where stages can be
evaluated in parallel include
\cite{MirankerLiniger67,EnenkelJackson97,ketcheson2014}.
Alternatively, one can use a predictor--corrector framework to
generate parallel-across-the-method time integrators.  This includes
parallel extrapolation methods \cite{Kappeller1996301}, and RIDC
integrators \cite{ChristliebMacdonaldOng2010,implicit-ridc}, which are
the focus of this paper.  A survey of parallel time integration
methods has recently appeared \cite{gander850}.

\subsection{Related Software}
\label{sec:related_software}
There are several well established software packages for solving
differential algebraic equations, however not many of them are able to
solve IVPs~\eqref{eqn:ivp} in parallel.  For sequential integrators,
probably the most well known are MATLAB routines {\tt ode45, ode23,
  ode15s} \cite{MR1719129} to solve their systems of differential
equations.  These schemes use embedded RK pairs or numerical
differentiation formulas (of the specified order) to approximate
solutions to the differential equations using adaptive time-stepping.
Readers might also be familiar with DASSL \cite{MR751605}, which
implements backward differentiation formulas of order one through
five.  The nonlinear system at each time-step is solved by Newton's
method, and the resulting linear systems are solved using routines
from LINPACK.  DASSL leverages the SLATEC Common Mathematical Library
\cite{vandevender1982slatec} for step-size adaptivity.  Also popular
are ODEPACK \cite{odepack} and VODE \cite{vode}, a collection of
fortran solvers for IVPs, SUNDIALS, a suite of robust time integrators
and nonlinear solvers \cite{sundials}, and there are a variety of ODE
and DAE time steppers implemented in PETSc \cite{petsc} and GSL
\cite{gsl}.

The selection of parallel solvers for IVPs is fairly
sparse.  
%The authors are aware of several maintained software
%implementations for solving IVPs in parallel.  
EPPEER \cite{EPEER} is a Fortran95/OpenMP implementation of explicit
parallel two-step peer methods \cite{Weiner2008} for the solution of
ODEs on multicore architectures.  PyPFASST \cite{PyPFASST} is a python
implementation of a modified parareal solver for ODEs and PDEs
\cite{emmett2012}.  XBRAID \cite{xbraid-package} is a C library that
implements a multigrid-reduction-in-time algorithm \cite{mgrit}, where
multiple time-grids of different granularity are distributed across
processors using MPI.  PFASST++ \cite{pfasst-package} is a C++
implementation of the `` parallel full approximation scheme in space
and time (PFASST) algorithm \cite{pfasst}.  There are other
implementations (such as the dependency-driven parareal framework
developed at Oakridge National Laboratory
\cite{Elwasif:2011:DFP:2132876.2132883}) that do not appear to be
available for download at present.

\section{Review of RIDC Methods}
\label{sec:methods}

Spectral deferred correction (SDC) \cite{Dgr00} provides an iterative
correction of an approximate solution by solving an integral
formulation of an error equation.  This integral form stabilizes the
classical differential deferred correction approach.  RIDC is a
re--formulation of SDC, pipelining successive calculations so that
corrections can be obtained in parallel with an appropriate time lag.
SDC, in contrast, is a sequential algorithm.
%High order
%approximations at $t_{n+1}$ are only obtained once \todo{fix: a high
%  order approximation is computed at $t_n$.}
Unlike the spectral deferred correction, which uses Gauss--Lobatto
nodes, RIDC uses uniformly spaced nodes to minimize the memory
footprint and to allow one to embed high order integrators
\cite{ChristliebOngQiu09a,ChristliebOngQiu09b}.

The basic idea of the IDC and RIDC approaches is to formulate associated error
IVPs which {\em correct} numerical errors from the solutions to
IVP~\eqref{eqn:ivp}; the parallelism arises from the ability to
simultaneously compute solutions to both IVP~\eqref{eqn:ivp} and
solutions to the associated error IVPs.  In this section, we review
the formulation of the error equations, discretizations, and parallel
properties of the RIDC algorithm.  Please refer to
\cite{ChristliebMacdonaldOng2010,implicit-ridc} for accuracy and stability
properties of the RIDC approach.

\subsection{Error IVPs}
Denote the (unknown) exact solution of IVP~\eqref{eqn:ivp} as $y(t)$,
and the approximate solution as $u(t)$, with $u(0)=y(0)$.  The error
in the approximate solution is $e(t) = y(t)-u(t)$.  Define the
residual (sometimes known as the defect) as $r(t) = u'(t)-f(t,u)$.
%$r(t) = u'(t)-f_S(t,u) - f_N(t,u)$. 
Then, the time derivative of the error satisfies
\begin{align*}
%  e'(t) = y'(t) - u'(t) = \left[f_S(t,u+e) + f_N(t,u+e)\right] 
%  - \left[f_S(t,u) + f_N(t,u) \right]- r(t).
  e'(t) = y'(t) - u'(t) = f(t,u+e) 
  - f(t,u) - r(t).
\end{align*}
Since $e(0)=u(0)-y(0)=0$, we have just derived the associated error
IVP. For stability, the integral form of the error IVP is preferred
\cite{Dgr00},
\begin{align}
%  \left(e + \int_0^t r(\tau)\,d\tau \right)' = 
%  \left[f_S(t,u+e) + f_N(t,u+e)\right] 
%  - \left[f_S(t,u) + f_N(t,u) \right].
  \left(e + \int_0^t r(\tau)\,d\tau \right)' = 
  f(t,u+e)
  - f(t,u).
  \label{eqn:error_ivp}
\end{align}

Observing that the corrected approximation $u+e$ is still an
approximation if the error equation~\eqref{eqn:error_ivp} is solved
numerically, we adopt a more general notation which will allow us to
iteratively correct the solution until a desired accuracy is reached.
Denote the initial approximation as $u^{[0]}$, the $p$th approximation
as $u^{[p]}$, and the error to $u^{[p]}$ as $e^{[p]}$.  Then, the
error equation can be rewritten as
\begin{align}
%  \left(e^{[p]} + \int_0^t r^{[p]}(\tau)\,d\tau \right)' = 
%  f_S(t,u^{[p]}+e^{[p]}) + f_N(t,u^{[p]}+e^{[p]}) 
%  - f_S(t,u^{[p]}) - f_N(t,u^{[p]}),
  \left(e^{[p]} + \int_0^t r^{[p]}(\tau)\,d\tau \right)' = 
  f(t,u^{[p]}+e^{[p]})
  - f(t,u^{[p]}),
  \label{eqn:error_ivp2}
\end{align}
where $r^{[p]} =  u^{[p]}(t)'-f(t,u^{[p]})$.
%$r^{[p]} =  u^{[p]}(t)'-f_S(t,u^{[p]}) - f_N(t,u^{[p]})$.

\subsection{Discretization}
% We consider three discretizations of the error
% IVP~\eqref{eqn:error_ivp}: (i) a forward Euler discretization if
% $f_S=0$, (ii) a backward Euler discretization if $f_N=0$, and (iii)
% a first order implicit--explicit (IMEX) method \cite{} in the
% general case.

With some algebra, 
%the following semi-discretizations ({\bf RH says: semi-discretizations of what?, why
%semi-discrete?} can be
%obtained. 
a first-order explicit discretization of (\ref{eqn:error_ivp2}), written in terms of the solution, gives
\begin{align}
  \label{eqn:error_fe}
  u_{n+1}^{[p+1]} = u_n^{[p+1]} 
  + \Delta t f(t_n,u_n^{[p+1]}) 
  - \Delta t f(t_n,u_n^{[p]})
  + \int_{t_n}^{t_{n+1}} f(\tau,u^{[p]})\,d\tau.
\end{align}
Likewise a first-order implicit discretization of (\ref{eqn:error_ivp2}) gives 
\begin{align}
  \label{eqn:error_be}
  u_{n+1}^{[p+1]} = u_n^{[p+1]} 
  + \Delta t f(t_{n+1},u_{n+1}^{[p+1]}) 
  - \Delta t f(t_{n+1},u_{n+1}^{[p]})
  + \int_{t_n}^{t_{n+1}} f(\tau,u^{[p]})\,d\tau.
\end{align}
In both semi-descretizations \eqref{eqn:error_fe} and
\eqref{eqn:error_be}, a sufficiently accurate quadrature is needed to
approximate the integrals present \cite{Dgr00}.  If a first
order predictor was applied to obtain an approximate solution to
\eqref{eqn:ivp}, and first order correctors such as
\eqref{eqn:error_fe} and \eqref{eqn:error_be} are used, approximating
the quadrature using
\begin{align*}
  \int_{t_n}^{t_{n+1}} f(\tau,u^{[p]})\,d\tau \approx
  \begin{cases}
    \displaystyle
    \sum_{\nu=0}^{p+1} \alpha_{p\nu} f(t_{n+1-\nu},u_{n+1-\nu}^{[p]}),
    &\mbox{ if } n \ge p,\\
    \displaystyle
    \sum_{\nu=0}^{p+1} \alpha_{p\nu} f(t_\nu,u_{\nu}^{[p]}), &\mbox{ if }  n<p,
  \end{cases},
\end{align*}
where $\alpha_{p\nu}$ are quadrature weights,
\begin{align*}
  \alpha_{p\nu} = 
  \begin{cases}
    \displaystyle
    \int_{t_n}^{t_{n+1}} \prod_{i=0, i\neq \nu}^{p+1}
    \frac{(t-t_{n+1-i})}{(t_{n+1-\nu}-t_{n+1-i})}\,dt,
    &\mbox{ if } n \ge p,\\
    \displaystyle
    \int_{t_n}^{t_{n+1}} \prod_{i=0, i\neq \nu}^{p+1}
    \frac{(t-t_{i})}{(t_{\nu}-t_{i})}\,dt,
    &\mbox{ if } n < p
  \end{cases}
\end{align*}
results in a $P$th order method, if $(P-1)$ such corrections are applied.

\subsection{Stability}
 A study of the (linear) stability of explicit RIDC methods is
 provided in \cite{ChristliebMacdonaldOng2010} and for implicit RIDC
 methods in \cite{implicit-ridc}.  The results indicate that the
 region of absolute stability of RIDC methods approach the region of
 absolute stability of the underlying predictor as the number of time
 steps increases.  Moreoever, for the implicit RIDC4-BE method
 preserves the $A$--stability property of backward Euler.

\subsection{Parallelization}
\begin{figure}[htbp]
  \begin{minipage}{0.7\textwidth}
    \centering
    \includegraphics{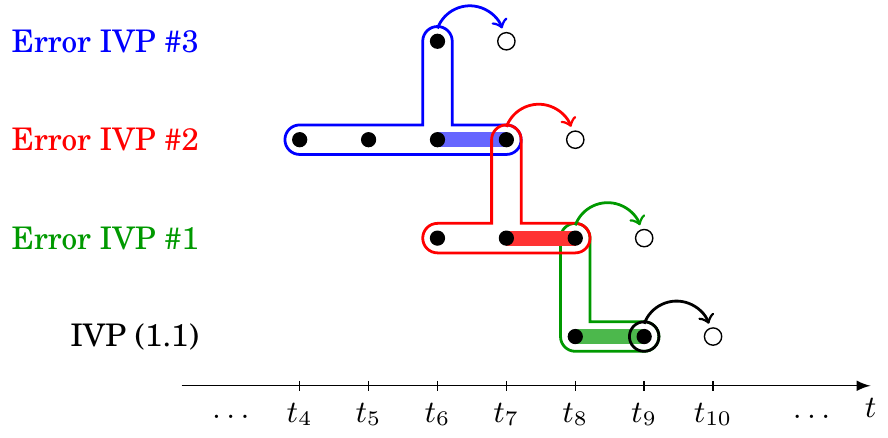}
  \end{minipage}\hfill
  \begin{minipage}{0.29\textwidth}
  \caption{In a RIDC method, solution values and correction terms are
    computed in a pipeline fashion.  For example, while a processor is
    computing a solution to IVP~\eqref{eqn:ivp} at $t_{10}$, a second
    processor computes corrections to the numerical error at time
    $t_9$, a third processor computes additional corrections at time
    $t_8$, and so forth.  (i.e., the open circles denote solutions
    that are simultaneously computed). The solid circles denote stored
    solution values that are needed for the quadrature approximation.}
  \label{fig:ridc}
  \end{minipage}
\end{figure}

%{\bf RH says: the authors instructions make particular mention of
%self-plagiarism, have to be careful about this}
As mentioned earlier, the parallelism arises from the ability to
simultaneously compute solutions to both IVP~\eqref{eqn:ivp} and
solutions to the associated error IVPs~\eqref{eqn:error_ivp2}.  This
is possible if there is some staggering to decouple solutions of
IVP~\eqref{eqn:ivp} and the error equations. As shown in
Figure~\ref{fig:ridc}, staggering of one timestep is required to
compute solutions in a pipeline parallel fashion.  For example, while
the predictor computes a solution at time $t_{10}$, the first
corrector computes the correction at time $t_9$, the second corrector
the second correction at time $t_8$, and so on.
We discuss the ``memory footprint'' and the startup routine
required by the RIDC method in Section~\ref{sec:memory_footprint}
before presenting a pseudo algorithm for the RIDC methods on page \pageref{alg:ridc_pseudo}.
% in
%Section~\ref{sec:ridc_pseudo}.
%Let $\mathcal{T}^{[p]}$

\subsection{Memory Footprint, Efficiency, Start-up and Shut-down}
\label{sec:memory_footprint}
Figure~\ref{fig:ridc} also shows  the ``memory footprint''
required to execute the RIDC method in a pipeline-parallel fashion. The
memory footprint are copies of the solution vector evaluated at
earlier correction/prediction levels and time steps; one can also
think of the memory footprint as the discretization stencil across the
different correction and prediction levels.  For a $P$th order RIDC
method, the $(P-1)$st correction update (i.e. solving error IVP
\#(P-1)) requires a stencil of size $(P+1)$, the $(P-2)$nd correction
requires an additional $(P-2)$ size stencil, the $(P-2)$nd correction
requires an additional $(P-3)$ size stencil, and so on.  The total memory
footprint required for a $P$th order RIDC method is
\begin{align*}
  \left(\sum_{i=1}^{P-1} (i+1) \right) + 1 = \frac{(P-1)(P)}{2} + (P-1) + 1
  =\frac{P(P+1)}{2}. 
\end{align*}

In \cite{ChristliebMacdonaldOng2010} it is shown that the ratio of
time steps taken by $P$th-order RIDC--Euler method, using $K$ steps
before a restart, to the number of steps taken by the forward Euler
method is
$$\gamma = 1+\frac{(P-1)^2}{K}.$$ This shows that the method becomes
more efficient (in terms of wall-clock time) as $K$ increases.  One
does have to balance a large value of $K$ with the possible increase
in error this may cause.  A study of this balance is provided in
\cite{ChristliebMacdonaldOng2010}.

\begin{figure}[htbp]
  \begin{minipage}{0.7\textwidth}
    \centering
    \includegraphics{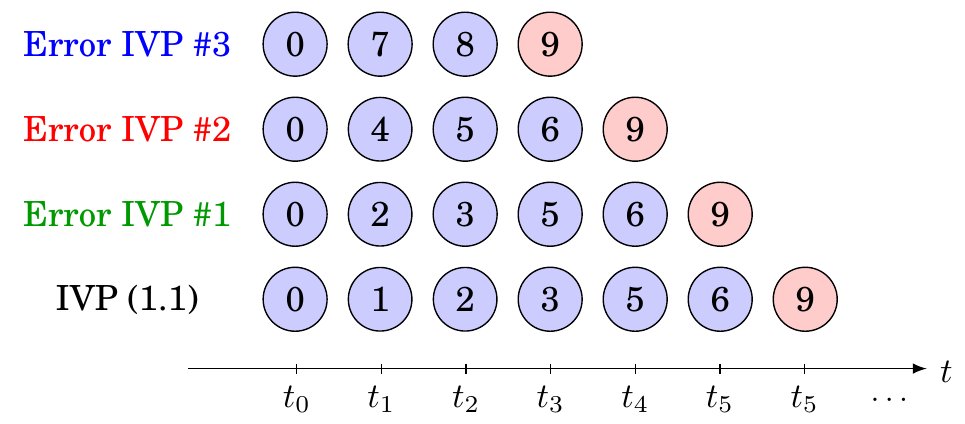}    
  \end{minipage}\hfill
  \begin{minipage}{0.29\textwidth}
  \caption{Start-up routine for a fourth-order RIDC method.  Observe
    that the predictor and lower order correctors are occasionally
    stalled to ensure that a minimal memory footprint is used for the
    RIDC method.  The fourth-order RIDC startup takes eight steps; from 
    step nine on, the RIDC method can be marched in a pipeline fashion.}
  \label{fig:ridc3_startup}
  \end{minipage}
\end{figure}

Because of the staggering, start-up steps are needed to fill the
memory footprint.  As discussed in \cite{ChristliebMacdonaldOng2010},
one should control the start-up steps to minimize the size of the
memory footprint; that is, it is more desirable to stall the
predictors and lower-level correctors initially (as appropriate) until
all predictors and correctors can be marched in a pipeline fashion
with the minimal memory footprint.  For example,
Figure~\ref{fig:ridc3_startup} shows the start-up routine for a
fourth-order RIDC method.  Initially, only the predictor advances the
solution from $t_0$ to $t_1$ in step one. In steps two and three, both
the predictor and first corrector are advanced to populate the memory
stencil in preparation for the second corrector.  In step four,
only the second corrector is advanced; the predictor and first
corrector are stalled because the memory stencil needed to advance the
second corrector from $t_1$ to $t_2$ is {\em the same} memory stencil
needed to advance the corrector from $t_0$ to $t_1$.

Although this concept is easy to grasp, the startup algorithm looks
non-intuitive at first glance.  Algorithm~\ref{alg:ridc_startup}
specifies the nuts-and-bolts of the start-up routine.  The RIDC method
can be run in a pipe-line fashion (with the minimal memory footprint)
after $startnum-1$ initialization steps, where $startnum =
min(1,\frac{p(p+1)}{2}-1)$.  For example, no initialization is
required if $p=1$.  If $p=4$, eight initialization steps are required
-- the RIDC method starts marching in a pipeline fashion at step nine.
\begin{algorithm}
  \caption{RIDC Startup-Routine}
  \label{alg:ridc_startup}
   $startnum = min(1,\frac{p(p+1)}{2}-1)$\;
  \For{$p=1$ \KwTo $(P-1)$}{
    march previous levels (i.e. $0,\ldots,(p-1)$) in a pipe for one step\;
    march current level $(p-1)$ steps\;
    march levels $0,\ldots,p$, in a pipe for one step\;
  }
\end{algorithm}
In the RIDC software, this startup routine is implemented using the
$filter$ variable.

The shut-down routine for the RIDC method is straightforward; each
predictor and corrector only advances the solution until the final
time, $t_F$, is reached.  The parallel RIDC pseudo-code is summarized
in Algorithm~\ref{alg:ridc_pseudo}.
\begin{algorithm}
\caption{RIDC Pseudo Code}
\label{alg:ridc_pseudo}
fill memory stencil, compute $startnum$ \;
\For{$nt = startnum$ \KwTo $NT$}{
  \ForPar{$p = 0$ \KwTo $(P-1)$}{
    \eIf{$p=0$} {
      use {\tt step} to advance solution on prediction level 
      (if $t_F$ not reached on prediction level)\;
    } {
      use {\tt corr\_fe} or {\tt corr\_be} to advance solution 
      on correction level $p$ 
      (if $t_F$ not reached  on correction level $p$)\;
    }
  }
  update memory stencil \;
}
\end{algorithm}

\section{RIDC Software}

To utilize popular sequential integrators as described in
Section~\ref{sec:related_software}, a user often specifies $f(t,y)$, the
range of integration $[0,T]$, the initial condition $y_0 = y(0)$ (and
for DASSL, the derivative $y'_0 = y'(0)$), and integrator parameters
(such as parameters for controlling step-size adaptivity). While these
general purpose time integration routines are convenient and easy to
use, this ``black-box'' approach (for example, a user does not have to
deal with the nonlinear solves arising from the backward
differentiation formulas) sometimes precludes the use of additional
information, such as the use of a problem-specific preconditioner,
sparsity of the matrices, or multigrid iterative solvers.

The RIDC software presented here differs from the type of
time-integration software mentioned above in that a first-order,
user-specified, advance for $t\to t+\Delta t$ is bootstrapped to
generate a high-order, parallel integrator using the integral deferred
correction framework described in Section~\ref{sec:methods}.

\subsection{Installation Instructions for Users}
The RIDC software is hosted at \url{http://mathgeek.us/software.html}.
Users should download the latest {\tt libridc-x.x.tar.gz}, and
uncompress that file using the command {\tt tar -zxvf
  libridc-x.x.tar.gz}.  

There are no prerequisites for building the base RIDC software and
examples in {\tt explicit/} and {\tt implicit/}.  To build the example
in {\tt brusselator\_gsl/}, the GNU Scientific Library \cite{gsl} and
headers need to be installed.  To build examples in {\tt
  implicit\_mkl/, brusselator\_mkl/} and {\tt
  brusselator\_radau\_mkl/}, the Intel Math Kernel Library needs to be
installed, and appropriate environment variables initialized.

In the top level directory {\tt ./configure --help} gives the possible
configuration options.  To configure using standard build options type
{\tt ./configure --prefix=/home/user/opt/libridc}. The library is
built by typing {\tt make \&\& make check \&\& make install}.  In this
instance the library and required header files would be installed in
{\tt home/user/opt/libridc/lib} and {\tt
  home/user/opt/libridc/include} respectively.  By default, only the
explicit and implicit examples are part of make check.  Users can
compile and check the MKL and GSL examples by typing {\tt
  --with-intel-mkl} or {\tt --with-gsl} in the configure step.

% For convenience, environment variables 

\subsection{Installation Instructions for Developers}

The development branch is hosted on GitHub, and can be obtained by
issuing the instructions: {\tt git clone
  https://github.com/ongbw/ridc}.  The git commit correlating to this
paper is {\tt f6c707e}.

The RIDC software is managed by the GNU build system.  As such, the
developer release requires GNU autoconf, automake, libtool, m4, make
and their respective prerequisites.  If there are version mismatches
between the RIDC software and the local system, issuing the commands
{\tt autoreconf -f} and {\tt automake -a -c} should resolve version
errors and warning.  To build the documentation, Doxygen must be
installed, as well as appropriate Doxygen pre-requisites.  For
example, to build a PDF manual documenting the source code, Doxygen
requires a \LaTeX compiler.

%% The developer version on github might require you
%% to install some updated packages, or you can type
%% {\tt autoreconf} and then {\tt  automake -a -c}
%% to customize to your system/libraries.  
%% To build a distribution file 
%% one types {\tt make dist}.
%% To build documentation, users have to have doxygen installed, and they
%% can enable their appropriate flavor, e.g.
%% {\tt ./configure --enable-doxygen-pdf} and then 
%% {\tt make \&\& make install \&\& make doxygen-doc}.
%% If no doxygen flavor is specified, it defaults to building html. 

\subsection{Running the Examples}
The directory {\tt examples/} includes five examples of utilizing the
RIDC library, and one example, {\tt examples/brusselator\_radau\_mkl}
that implements a three stage, fifth-order Radau method to provide a
basis of comparison with the RIDC integrators.  Depending on the
options selected in the {\tt ./configure} step, some or all of these
examples are built and run during during the {\tt make check} process.
Alternatively, a user can compile and run an example seperately after
the {\tt./configure} step.  For example, the subdirectory {\tt
  examples/explicit/} contains the code to solve
$$y_i' = -(i+1)ty_i,\,\, y_i(0)=1,\quad i=1,2,$$ using RIDC with an
explicit Euler step function.  To compile this specific example, move
into the {\tt examples/explicit} subdirectory and type {\tt make
  explicit}.  The executable {\tt explicit} takes as input the order
required and the number of time steps. For example {\tt
  ./explicit\_ode1.exe 4 100} solves the system of ODEs using fourth
order RIDC with 100 time steps.  A shell script {\tt run.sh} is
provided to run the RIDC integrator with different numbers of time
steps for a convergence study.  A simple matlab or octave script {\tt
  convergence.m} is included to test the order of convergence.  {\tt
  octave convergence.m} gives the slope and intercept for the linear
fit of log of the error versus log of the time step.  In this example
we obtain a slope of $-4.0630$ indicating the we indeed have an order
4 method.

\subsection{Using the RIDC library}

To utilize the RIDC library, a main program should specify problem
parameters (using the PARAMETER structure), initial values and the
order of the RIDC integrator desired.  The solution is integrated
using a call to the {\tt ridc\_fe} or {\tt ridc\_be} functions The
user also needs to specify template functions for the right hand side
of the ODE $y'=f(t,y)$ and a step function which advances the
function from $t_n$ to $t_{n+1}$.
% In the file {\tt driver.cpp} the user specifies the PARAMETER
%structure -- which contains the problem %descriptors including the
%initial and final times, the time step, the dimension of the problem
%and so on.  The initial condition of the system is prescribed here
%and the solution is integrated using a call to the {\tt ridc\_fe} or
%{\tt ridc\_be} functions.  This would also be the natural place to
%control solution output.  
This step routine may be complicated requiring large scale linear
algebra provided by external external libraries or possibly a
nonlinear solve.  The {\tt examples/brusselator\_gsl} directory
contains such an example.  This example uses a backward Euler step for
a nonlinear system of ODEs.  The step function uses a Newton iteration
(see the functions {\tt newt} and {\tt jac}) and the GNU scientific
library (GSL) \cite{gsl} to solve for the Newton step.  The functions
{\tt newt} and {\tt jac} required by step are defined and declared in
{\tt brusselator.cpp}.  To link against the RIDC library, include the
following arguments to the compilation command:
\begin{center}
  {\tt -L/home/user/opt/libridc/lib -I/home/user/opt/libridc/include -lridc}
\end{center}
For information about the RIDC functions, please refer to the
documentation located in {\tt doxygen-doc/} or online at
\url{http://mathgeek.us/software.html}.

%% The files {\tt driver.cpp} and {\tt
%%   explicit.cpp} are then compiled and linked against the {\tt ridc}
%% library located in {\tt /home/user/opt/libridc}.

%% The Makefile will need to be edited as well.  In particular any
%% required libraries required by step must be linked here.  For example
%% {\tt -lgsl} and {\tt -lm} are included in the {\tt Makefile} for the 
%% brusselator--gsl example. 

\subsection{Under the hood}
\begin{figure}[htbp]
  \begin{minipage}{0.6\textwidth}
    \centering
    \includegraphics{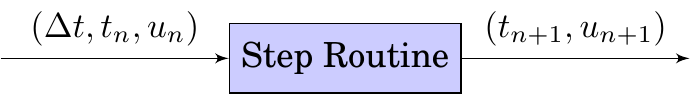}
  \end{minipage}\hfill
  \begin{minipage}{0.39\textwidth}
  \caption{User-defined step routine that advances a solution from
$t_n$ to $t_{n+1}$.
      }
  \label{fig:step}
  \end{minipage}
\end{figure}

The RIDC software and examples are coded in C++; task parallelism is
achieved using OpenMP threads to solve the predictors and the
correctors in parallel.  This mode of parallelism was chosen to
accommodate the data movement/communication required by the RIDC
algorithm when solving equations~\eqref{eqn:error_fe} and
\eqref{eqn:error_be}.  We assume that the user-defined step routine to
advance the solution is a first-order sequential integrator, although
with some minor modifications to the RIDC software provided,
bootstrapping higher order integrators is possible.  The RIDC
software can also be modified to leverage a thread-safe user-defined
step routine, for example a CUDA-accelerated step routine
\cite{ridc-gpgpu} or an MPI-parallelized step routine
\cite{OngHaynesChristlieb2012} can be utilized, see Section
\ref{sec:gener}.  If the step routine uses an explicit Euler
integrator, the RIDC software assumes that $u_{n+1}$ satisfies
\begin{align*}
  u_{n+1} - u_n = \Delta t f(t_n,u_{n}).
\end{align*}
If the step routine uses an implicit Euler integrator, the RIDC
software assumes that $u_{n+1}$ satisfies
\begin{align*}
  u_{n+1} - u_n = \Delta t f(t_{n+1},u_{n+1}).
\end{align*}
The RIDC software treats this step routine as a black box, as depicted
in Figure~\ref{fig:step}.

\begin{figure}[htbp]
  \begin{minipage}{0.6\textwidth}
    \centering
    \includegraphics{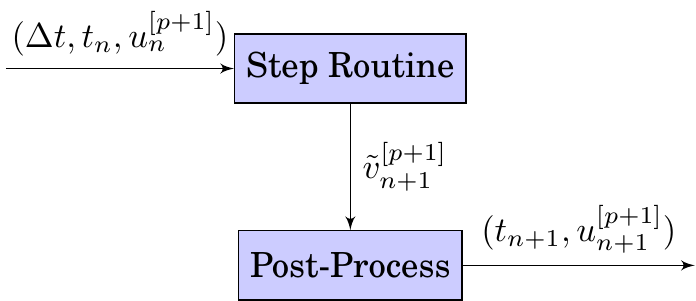}    
  \end{minipage}\hfill
  \begin{minipage}{0.39\textwidth}
  \caption{A visualization of the RIDC wrapper to obtain a solution to
    equation~\eqref{eqn:error_fe}.  The post process takes an input
    $\tilde{v}_{n+1}^{[p+1]}$ and returns $\tilde{v}_{n+1}^{[p+1]}- \Delta t
    f(t_n,u_n^{[p]}) + \int_{t_n}^{t_{n+1}} f(\tau,u^{[p]})\,d\tau$.}
  \label{fig:ridc_fe}
  \end{minipage}
\end{figure}

The RIDC functions solve equations~\eqref{eqn:error_fe} and
\eqref{eqn:error_be} by creating the necessary data structures to
store copies of the solution vector described in
Section~\ref{sec:memory_footprint}, and then performing the
appropriate algebraic computations on these stored solution values.
First, consider the explicit Euler discretization of the error
equation~\eqref{eqn:error_fe}.  Observe that $u_{n+1}^{[p+1]}$ can be
constructed by applying the user-defined step routine to
$u_{n}^{[p+1]}$ to obtain $\tilde{v}_{n+1}^{[p+1]}$, and then adding $
- \Delta t f(t_n,u_n^{[p]}) + \int_{t_n}^{t_{n+1}}
f(\tau,u^{[p]})\,d\tau$ to $\tilde{v}_{n+1}^{[p+1]}$ to finally obtain
$u_{n+1}^{[p+1]}$.  The explicit RIDC wrapper is displayed in
Figure~\ref{fig:ridc_fe}.

\begin{figure}[htbp]
  \begin{minipage}{0.6\textwidth}
    \centering
    \includegraphics{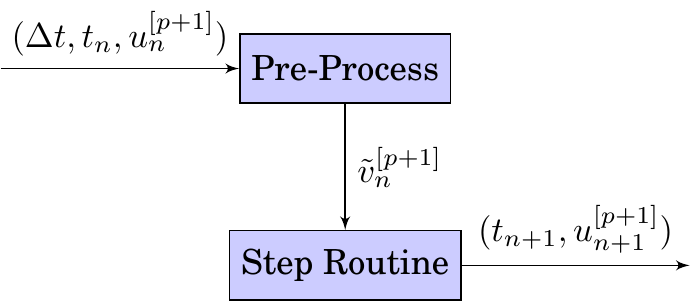}    
  \end{minipage}\hfill
  \begin{minipage}{0.39\textwidth}
  \caption{A visualization of the RIDC wrapper to obtain a solution to
    equation~\eqref{eqn:error_be}.  The pre-process takes an input
    $u_n^{[p+1]}$ and returns $\tilde{u}_{n}^{[p+1]}- \Delta t
    f(t_n,u_n^{[p+1]}) + \int_{t_n}^{t_{n+1}} f(\tau,u^{[p]})\,d\tau$.}
  \label{fig:ridc_be}
  \end{minipage}
\end{figure}

A similar observation can be made about the implicit Euler
discretization of the error equation~\eqref{eqn:error_be}, however,
one first constructs the intermediate value $\tilde{v}_{n}^{[p+1]} =
u_n^{[p+1]} - \Delta t f(t_n,u_n^{[p]}) + \int_{t_n}^{t_{n+1}}
f(\tau,u^{[p]})\,d\tau$, and then applies the user-defined step
function to $\tilde{v}_{n}^{[p+1]}$.  The implicit RIDC wrapper is
displayed in Figure~\ref{fig:ridc_be}.
%{\bf RH says: it looks like $\tilde{v}_{n+1}^{[p+1]}$ is incorrect, shouldn't it be
%$u_n^{[p+1]}-\Delta t f(t_{n+1},u_{n+1}^{[p]})+\int$?}
\subsection{Discussion}
%% \label{sec:discussion}
%% \todo{
%% \begin{itemize}
%% \item memory requirements
%% \item overhead (when can we expect parallel speedup?)
%% \end{itemize}
%% }

The computational overhead of RIDC methods resides mainly in the
quadrature approximation, and the subsequent linear combinations used
to compute the corrected solutions. Provided this computational
overhead is small compared to an evaluation of the step routine, good
parallel speedup is achieved.  In practice, this is almost always the
case for implicit RIDC methods where solutions to linear equations,
and/or Newton iterations are required.  For explicit RIDC methods,
good parallel speedup is only observed when the step routine is
sufficiently expensive, such as in the computation of self-consistent
forces for an $n$-body problem \cite{ChristliebMacdonaldOng2010}.

As mentioned in Section~\ref{sec:memory_footprint}, the RIDC method
has to store copies of the solution vector evaluated at ealier
correction/prediction levels.  Although this memory requirement might
appear restrictive, the memory footprint for high order single-step,
multi-step or general linear methods are similar.  Implicit RIDC
methods also benefit from the loose coupling between the prediciton
and corection equations; whereas a general implicit $s$-stage implicit
RK method neccessitates the solution of a system of (potentially
nonlinear) $sN$ equations, where $N$ are the number of differential
algebraic equations.  A $p$th-order RIDC method constructed using
backward Euler integrators requires the solution of $p$ decoupled
systems of $N$ (potentially nonlinear) algebraic equations.

\subsection{Possible Generalizations}
\label{sec:gener}
For clarity, only the simplest variant of the RIDC method
(constructed using first order Euler integrators, uniform
time-stepping, serial computation of the step routine) has been
presented, and released as part of the base software version.  Here,
we make some remarks on how the base version of the software can be
modified by the user to accommodate several generalizations discussed
in this secton; indeed, the authors will release (when possible)
modified versions of the software within the source repository that
illustrate how to generate generalized RIDC integrators.

{\em Step-size adaptivity for error control}: In \cite{adaptive-ridc},
various variants of adaptive RIDC methods were presented.  In the
simplest variant, one uses standard error control stratagies to
adaptively select step-sizes while solving IVP~\eqref{eqn:ivp}.  These
adaptively selected step-sizes are used for solving the error
equations~\eqref{eqn:error_ivp}.  To build step-size adaptivity into
the provided RIDC software, the following modifications will be
needed: (i) modify the time-loop appropriately to allow for
non-uniform steps, (ii) modify the driver file appropriately to take a
user-defined tolerance (as opposed to the number of time steps), (iii)
recompute the integration matrix containing the quadrature weights at
every time step.  The user will presumably provide an additional {\tt
  adapt\_step} function, which takes as inputs the solution at time
$t$, the previous time step used, $\Delta t_{old}$, a tolerance $tol$,
and returns the time step selected, $\Delta t$, and the solution at
the new time step, $t + \Delta t$.

{\em Restarts}: As discussed in \cite{ChristliebMacdonaldOng2010},
the RIDC method accumulates error while running in a pipeline fashion
-- the most accurate solution does not propagate to the earlier
prediction/correction levels.  In some cases, it might be advantageous
to stop the RIDC method, and use the most accurate solution to
``restart'' the computation.  This requires only a simple modification
to the main RIDC loop in {\tt ridc.cpp}.

{\em Constructing RIDC methods using higher-order integrators}: With a
few modifications, it is possible to use higher-order single step
integrators within the RIDC software.  The memory stencil, integration
matrix and quadrature approximations will need to be modified in {\tt
  ridc.cpp}.

{\em Semi-implicit RIDC methods}: Although semi-implicit RIDC methods
have been constructed and studied in \cite{ridc-gpgpu}, it is in
general not possible to wrap a user-defined semi-implicit step
function to solve the error equation~\eqref{eqn:error_ivp}.  
Consider the IVP
\begin{align*}
  {\bf y}'(t) = f_N(t,{\bf y}) + f_S(t,{\bf y}),  
\end{align*}
where $f_S$ contains stiff terms and $f_N$ contains the nonstiff terms.
A first-order user-defined step function to solve this IVP would look like
%{\bf RH says: what are $f_N$ and $f_S$ ??} satisfies
\begin{align*}
  u_{n+1} - u_n
  = \Delta t f_N(t_n,u_n) 
  + \Delta t f_S(t_{n+1},u_{n+1}),
\end{align*}
whereas the first-order IMEX discretization of the error
equation~\eqref{eqn:error_ivp} is
\begin{align*}
  u_{n+1}^{[p+1]} = u_n^{[p+1]} 
  + \Delta t \left[
     f_N(t_n,u_n^{[p+1]}) + 
    f_S(t_{n+1},u_{n+1}^{[p+1]}) 
    \right]
  - \Delta t \left[
    f_N(t_n,u_n^{[p]}) +
    f_S(t_{n+1},u_{n+1}^{[p]})
    \right]\\
  + \int_{t_n}^{t_{n+1}} \left[ f_N(\tau,u^{[p]}) +f_S(\tau,u^{[p]})
    \right]\,d\tau.
\end{align*}
Althought it is not obvious how to automaticaly bootstrap a semi--implicit step function,  a user 
can leverage the data structures and quadrature
approximations in {\tt ridc.cpp} to construct a new {\tt corr\_fbe}
function, which should look similar in structure to the users' {\tt
  step} function.

{\em Using accelerators for the step routine}: Many computing clusters
feature nodes with multiple accelerators, e.g. Nvidia GPGPUs or Intel
Xeon Phis.  If the user wishes to provide a step routine that is
accelerated using these emerging architectures, the RIDC code can be
modified to leverage {\em multiple} accelerators in a computational
node.  Modifications that are required include: adding an input
variable ``level'' (an integer from 0 to $p-1$, where $p$ is the
desired order / number of accelerators in the system) into the step
routine, a function call within the step function to specify the
appropriate accelerator, e.g. {\tt cudaSetDevice} for the NVIDIA
GPGPUs, and a modification of {\tt ridc.cpp} so that the
prediction/correction level is fed into the step function, ensuring
that the linear algebra is performed on the appropriate accelerator.

{\em Using distributed MPI for the step routine}: Although the RIDC
software can be modified to allow for an MPI-distributed step routine
(provided this step-routine is thread safe), we showed in
\cite{ridc-dd-implementation} that a tighter coupling of the hybrid
MPI-OpenMP formulation to reduce the number of messages is necessary
for performance.

\section{Numerical Experiment}

\begin{figure}[htbp]
  \begin{minipage}{0.7\textwidth}
    \centering
    \includegraphics{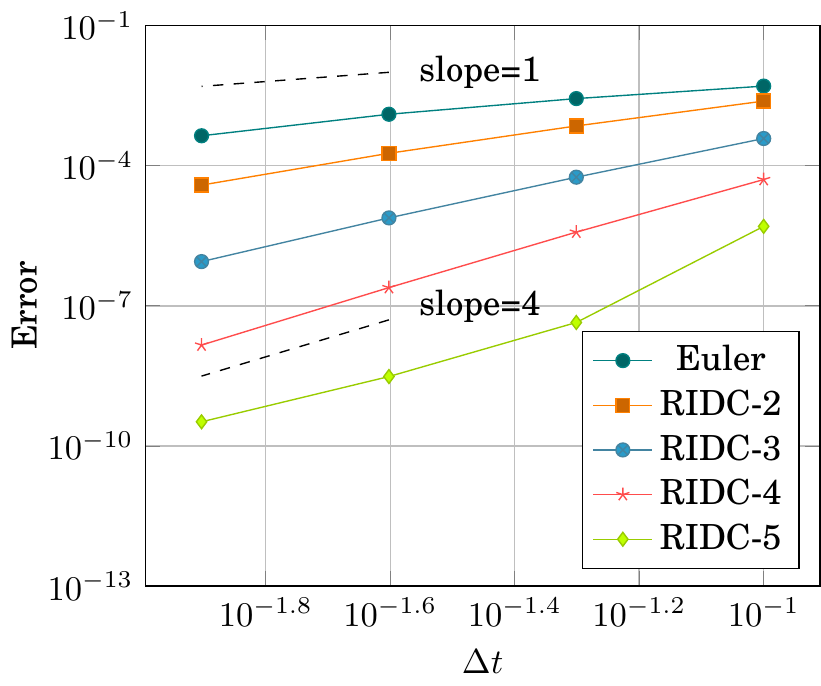}    
  \end{minipage}\hfill
  \begin{minipage}{0.29\textwidth}
    \caption{\label{fig:bruss_conv} Standard convergence study of
      error versus time step, $\Delta t$, showing that RIDC methods
      achieve their designed orders of accuracy.}
  \end{minipage}
\end{figure}

The software includes several examples verifying that the RIDC methods
attain their designed orders of accuracy.  As mentioned, these examples
also serve as templates for the user to bootstrap their own first order
time integration methods to give a parallel--in--time approximation.  
Good parallel speedup is
observed when the computational overhead for the RIDC methods (namely,
the quadrature approximation and the linear combinations to compute
the corrected solutions) is small compared to an evaluation of the
step routine.  Here, we present the numerical results for the Brusselator
in $\mathbb{R}^1$.
\begin{align}
  \label{eqn:bruss_1d}
  u_t &= A + u^2 v - (B+1) u + \alpha u_{xx},\\
  \nonumber
  v_t &= Bu - u^2v + \alpha v_{xx},
\end{align}
with $A=1$, $B=3$ and $\alpha=0.02$, initial conditions
\begin{align*}
  u(0,x) = 1 + \sin(2\pi x), \qquad  v(0,x) = 3,
\end{align*}  
and boundary conditions
\begin{align*}
  u(t,0) = u(t,1) = 1, \qquad
  v(t,0) = v(t,1) = 3.
\end{align*}
A central finite difference approximation is used to discretize
equation~\eqref{eqn:bruss_1d}.  The resulting nonlinear system of
equations is solved using a Newton iteration.  In the timing results,
the Intel Math Kernel Library (MKL) is used to solve the linear system
arising in each Newton iteration.  The code for this example can be
found in the {\tt examples/brusselator\_mkl}
directory. Figure~\ref{fig:bruss_conv} shows a standard convergence
study of error versus number of timesteps to demonstrate that the RIDC
software bootstraps the first order integrator to generate a
high-order method of the desired accuracy.

\begin{figure}[htbp]
  \centering
  \includegraphics{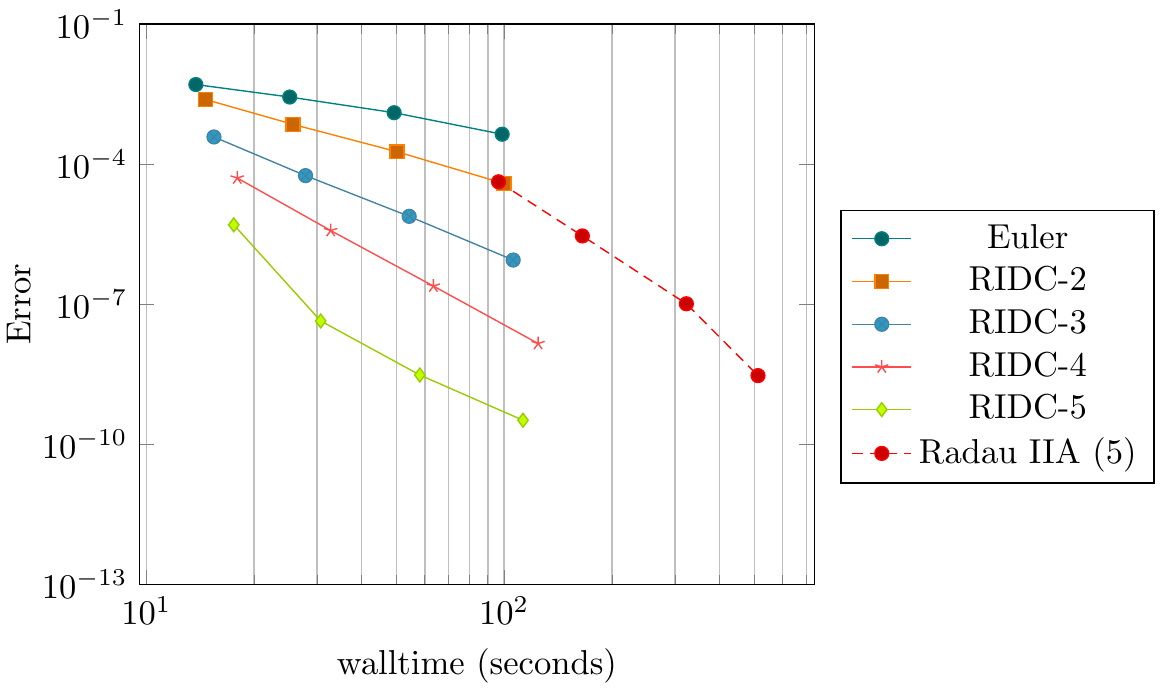}    
    \caption{ \label{fig:bruss_weak_scaling} The error as a function
      of walltime is plotted for various RIDC methods.  Here, two
      computing cores (set via {\tt OMP\_NUM\_THREADS=2}) is used to
      compute the second order RIDC method (RIDC-2), three computing
      cores are used to compute RIDC-3, four computing cores are used
      to compute RIDC-4, and give compute cores are used to compute
      RIDC-5.  A single computing core was used to compute Radau IIA.
      The RIDC software computes a $p$th order solution in
      approximately the same wall clock time as an Euler solution,
      provided $p$ computing cores are available.  The parallel RIDC
      methods also provide good speedup over a serial Radau IIA integrator.}
\end{figure}

In Figure~\ref{fig:bruss_weak_scaling}, the walltime used to compute
each ridc method is plotted to show the ``weak scaling'' capability of
RIDC methods.  For example, the fourth-order RIDC method computes a
solution using four computing cores that is 3-5 orders of magnitude
more accurate than the first order Euler solution in approximately the
same wallclock time.  Tming results using a serial three-stage,
fifth-order RADAU IIA integrator is also presented.  A fifth order
RIDC method (with five computing cores) provides a solution with
comparable accuracy in 10\% of the walltime.  The scaling studies were
performed on a single computational node consisting of a dual socket
Intel E5-2670v2 chipset.

%% Lastly, Figure~\ref{fig:bruss_strong_scaling} shows the results of a
%% strong scaling study, where the number of computing cores used in an
%% eighth-order RIDC method is varied.  Good small-scale parallel speedup
%% is observed.
%% \begin{figure}[htbp]
%%   \caption{todo}
%%   \label{fig:bruss_strong_scaling}
%% \end{figure}

\section{Conclusions}
In this paper, we presented the revisionist integral deferred
correction (RIDC) software for solving systems of initial values
problems.  The approach bootstraps lower order time integrators to
provide high order approximations in approximately the same wall clock
time, providing a multiplicative increase in the number of compute
cores utilized.  The C++ framework produces a parallel--in--time
solution of a system of initial value problems given user supplied
code for the right hand side of the system and the sequential code for
a first-order time step.  The user supplied time step routine may be
explicit or implicit and may make use of any auxiliary libraries which
take care of the solution of the nonlinear algebraic systems which
arise or the numerical linear algebra required.

% Bibliography
\bibliographystyle{ACM-Reference-Format-Journals}
\bibliography{ridc}

\end{document}